\pdfoutput=1
\documentclass[aps, twocolumn, floatfix, superscriptaddress, prl]{revtex4}
\usepackage{graphicx}
\DeclareGraphicsExtensions{.pdf}
\usepackage{amsmath,amssymb,bbold,bm,color}
\usepackage{float}
\usepackage{epstopdf}
\usepackage{tabularx}
\usepackage{braket}
\usepackage{booktabs}
\usepackage{array}
\usepackage{blindtext}
\newcolumntype{L}[1]{>{\raggedright\let\newline\\\arraybackslash\hspace{0pt}}m{#1}}
\newcolumntype{C}[1]{>{\centering\let\newline\\\arraybackslash\hspace{0pt}}m{#1}}
\newcolumntype{R}[1]{>{\raggedleft\let\newline\\\arraybackslash\hspace{0pt}}m{#1}}
\usepackage{calrsfs}
\DeclareMathAlphabet\mathbfcal{OMS}{cmsy}{b}{n}
\DeclareMathAlphabet{\pazocal}{OMS}{zplm}{m}{n}

\begin{document}

\title{Twist-induced magnon Landau levels in honeycomb magnets}

\author{Tianyu Liu}
\affiliation{Max-Planck-Institut f\"ur Physik komplexer Systeme, 01187 Dresden, Germany }

\author{Zheng Shi}
\affiliation{Dahlem Center for Complex Quantum Systems and Physics Department, Freie Universit\"at Berlin, 14195 Berlin, Germany}

\begin{abstract} 
Lattice deformation resulting from elastic strain is known to spatially modulate the wave function overlap of the atoms on the lattice and can drastically alter the properties of the quasiparticles. Here we elaborate that a twist lattice deformation in two-dimensional honeycomb quantum magnet nanoribbons is equivalent to an elastic gauge field giving rise to magnon Landau quantization. When the ground state is ferromagnetic, dispersive Dirac-Landau levels are induced in the center of magnon bands, while for antiferromagnetic nanoribbons, the twist results in dispersive equidistant Landau levels at the top of magnon bands. The dispersions for both types of Landau levels are derived in the framework of the band theory.
\end{abstract}

\date{\today}
\maketitle

\textit{Introduction}.--Strain engineering is a powerful tool in tuning properties of quantum matter, such as spin transport \cite{huangbing2017, csahin2019}, thermal conductivity \cite{meng2019, seijas2019}, and quantum anomaly \cite{cifuentes2016}. In particular, twisting one layer of bilayer graphene with respect to the other by certain ``magic'' angles \cite{bistritzer2011} results in spatial modulation of electron tunneling between the layers and produces flat ``Moir\'e bands'' responsible for the correlated insulating phase \cite{cao2018a} and the unconventional superconductivity \cite{cao2018b}. Properly tuned strain can close or open band gaps in topological quantum matter and induce phase transitions between distinct topological phases \cite{zhu2016, shao2017, guan2017, owerre2018, zhang2019, mutch2019}. 

Perhaps the most investigated and best understood strain effects are those associated with Dirac materials, where strain is famously equivalent to an elastic gauge field \cite{guinea2010a, levy2010, vozmediano2010, rechtsman2013, cortijo2015, pikulin2016, grushin2016, cortijo2016, sumiyoshi2016, arjona2017, liu2017a, liu2017b, massarelli2017, matsushita2018, kobayashi2018, nica2018, ferreiros2018, liu2019, liu2020}. A circular bend in 3D Dirac/Weyl semimetals and superconductors induces a uniform pseudo-magnetic field giving rise to Dirac-Landau levels \cite{arjona2017, liu2017a, liu2017b}. A uniform elastic gauge field can also be generated by twisting 3D Weyl systems around the axis on which Weyl points are located \cite{pikulin2016, matsushita2018, kobayashi2018, liu2019}. Though first theoretically proposed \cite{guinea2010a} and experimentally discovered \cite{levy2010} in graphene, the strain-induced gauge field in graphene and other 2D Dirac materials is not uniform for simple lattice deformations such as bend \cite{guinea2010b, chang2012, stuij2015} or twist \cite{zhang2014}, causing some difficulty in obtaining insights of the strain-induced Landau levels  (LLs).

In this Letter, we propose a simple strategy in the framework of band theory to obtain the dispersion of twist-induced LLs at Brillouin zone (BZ) corners for both ferromagnetic (FM) and antiferromagnetic (AF) honeycomb nanoribbons, whose magnon bands in the absence of twist exhibit Dirac cones and quadratic peaks, respectively. We demonstrate that the effect of twist is to relocate the Dirac cones (quadratic peaks) such that the dispersive LLs of ferromagnets (antiferromagnets) can be understood as pulled out by the displaced Dirac cones (quadratic peaks) from those at the BZ corners. Based on this observation, we show that a correspondence between the crystal momentum of the nanoribbon and the twist-induced elastic gauge field can be drawn to explicitly give the momentum dependence of the twist-induced LLs for both ferromagnets and antiferromagnets.

\textit{Twisted Heisenberg model}.--We consider a Heisenberg model defined on a honeycomb lattice of spins with only nearest-neighbor interactions
\begin{equation} \label{Heisenberg}
H=\sum_{\bm r} \sum_{i=1}^3 J_i \bm S_A(\bm r) \cdot \bm S_B (\bm r + \bm \alpha_i),
\end{equation}
where $\bm r$ denotes the position of a generic lattice site belonging to the $A$ sublattice and vectors $(\bm \alpha_1, \bm \alpha_2, \bm \alpha_3) = (\frac{\sqrt 3}{2} a\hat x + \frac{1}{2} a\hat y, -\frac{\sqrt 3}{2} a\hat x + \frac{1}{2} a\hat y, -a\hat y)$, with $a$ being the nearest-neighbor distance, connect this site to its three nearest-neighboring sites on the $B$ sublattice. $J_i=J(\alpha_i)$ is the interaction strength between the spin-$S$ located at $\bm r$ and its $i$-th nearest neighbor at $\bm r + \bm \alpha_i$. In this Letter, we will assume isotropic interaction $J(\alpha_i)=J$ for transparency. At sufficiently low temperature, the honeycomb magnet exhibits FM (AF) order when $J<0$ ($J>0$).

\begin{figure}
\includegraphics[width = 8.6cm]{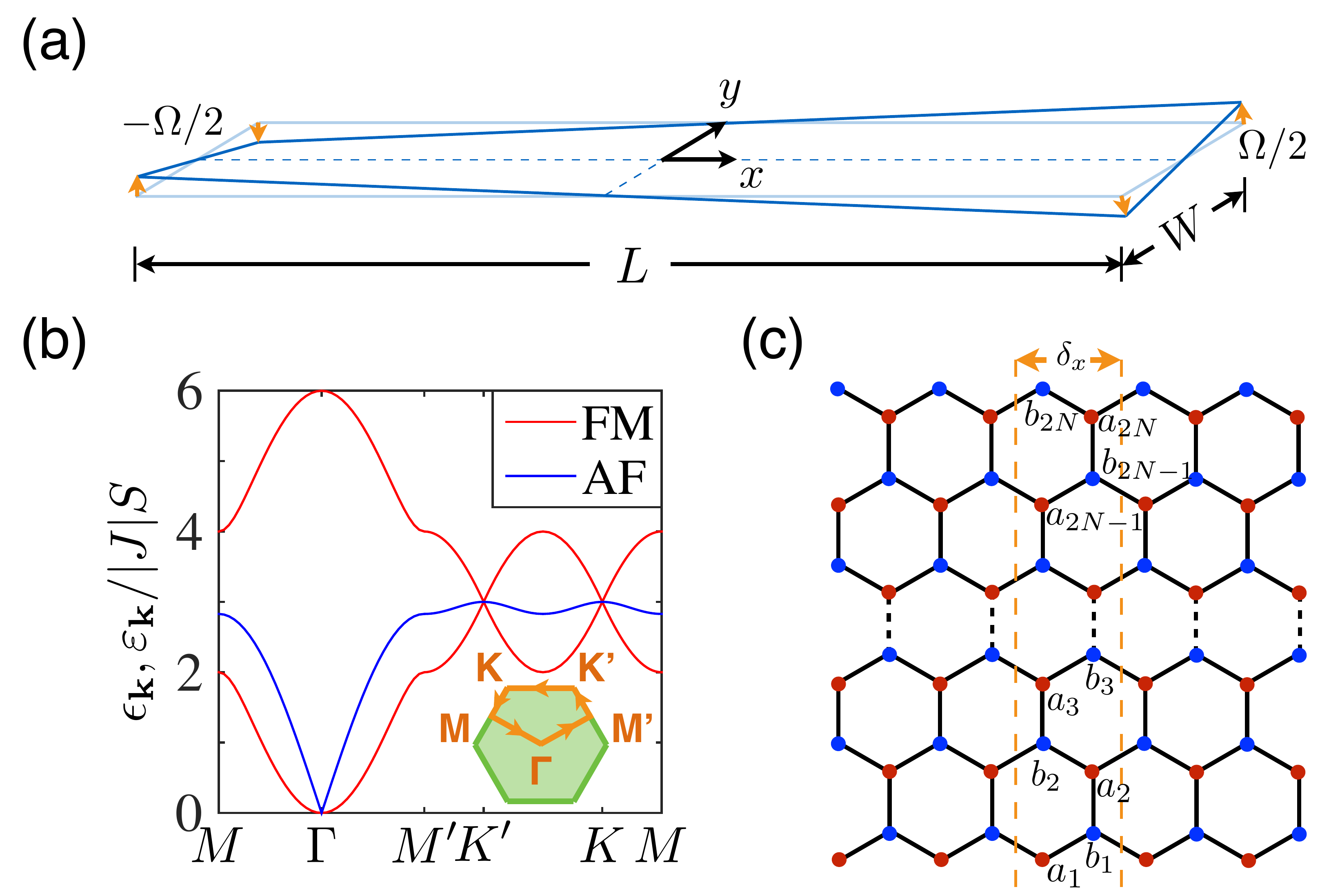}
\caption{(a) A twisted honeycomb magnet nanoribbon (blue) is obtained by applying torsional strain to a normal nanoribbon (light blue) of length $L$ and width $W$ to rotate the right (left) edge around the $x$ axis by a small angle $\Omega/2$ ($-\Omega/2$). (b) The dispersion of the honeycomb ferromagnet (red) and antiferromagnet (blue) along the path connecting the high-symmetry points of the Brillouin zone (Inset). (c) A zigzag nanoribbon with $2N$ $A$ and $B$ sublattice sites in a unit cell marked by two orange dashed lines. The width of the unit cell is $\delta_x=\sqrt{3}a$.} \label{fig1}
\end{figure}

In the presence of lattice deformation, the most important effect can be incorporated to the Heisenberg model (Eq.~\ref{Heisenberg}) by amending the nearest-neighbor interaction to $J_i = J \exp(-\beta \frac{\delta_i-\alpha_i}{\alpha_i})$, where $\delta_i$ is the bond length associated with the $i$-th nearest neighbor after the deformation and $\beta$ is the Gr\"uneisen parameter of order unity \cite{ferreiros2018, nayga2019}. Without loss of generality, we take $\beta=1$ for the following analytical derivations and numerical simulations. In particular, for the twist deformation illustrated in Fig.~\ref{fig1}(a), a nanoribbon is twisted around $x$ axis in such a way that a lattice site originally located at position $\bm r = (x, y, 0)$ is relocated to $\bm r + \bm u(\bm r) = (x, y \cos \lambda x, y\sin \lambda x)$, where $\lambda=\Omega/L$ measures the rotational angle of a $y$-direction chain per unit length along the $x$ direction. Consequently, the bond length after the twist becomes $\delta_i = [\alpha_i^2+\lambda^2\alpha_{i,x}^2(y^2+y\alpha_{i,y})]^{1/2}$. The resulting nearest-neighbor interactions are then the exponentially decaying $J_1=J_2=J\exp \{1-[1+\frac{3}{4}\lambda^2(y^2+\frac{a}{2}y)]^{1/2}\}$ and $J_3=J$. Specifically, for a narrow nanoribbon with sufficiently small twist, the bond length can be estimated as $\delta_i = \alpha_i[1+\lambda^2 \alpha_{i,x}^2 (y^2+y\alpha_{i,y})/2\alpha_i^2]$, giving rise to nearest-neighbor interactions 
\begin{equation} \label{quad}
J_1 = J_2 = J - \tfrac{3}{8} \lambda^2 (y^2+\tfrac{a}{2}y) J, \quad J_3=J.
\end{equation}

\begin{figure}
\includegraphics[width = 8.6cm]{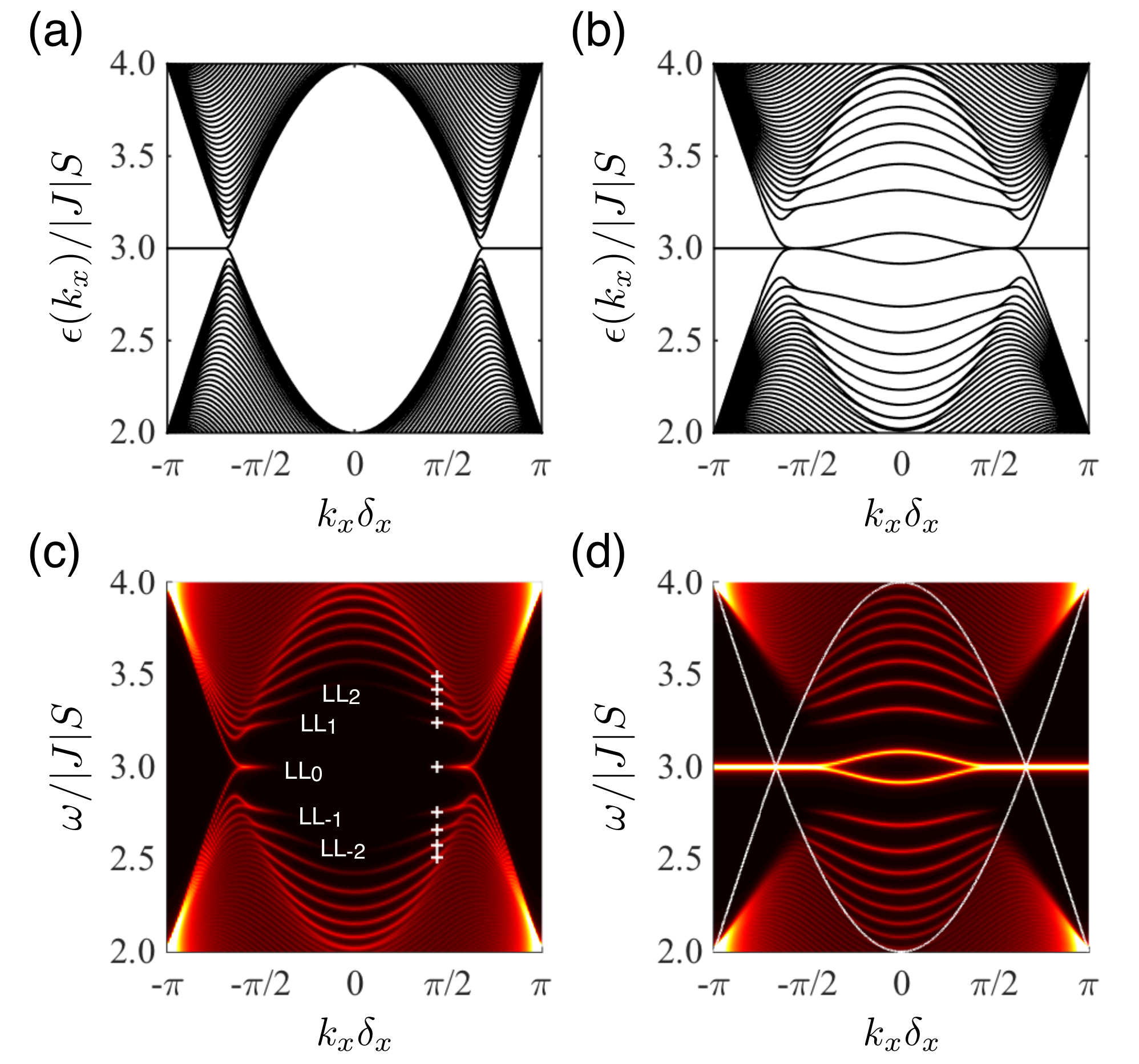}
\caption{Spectral properties of a zigzag FM nanoribbon with $2N=80$ $A$ and $B$ sublattice sites. (a) The spectrum of an untwisted nanoribbon exhibiting two Dirac cones connected by flat edge states. (b) The spectrum of a twisted nanoribbon with $\lambda a =0.032$. (c) The bulk spectral function of (b) with dispersive Dirac-Landau levels in the vicinity of $K$ valley labelled. The crosses mark the positions of LLs at the selected momentum $k_x\delta_x=1.382$. (d) The edge spectral function of (b) with the white curves representing the envelopes of the Dirac cones.} \label{fig2}
\end{figure}

\textit{Honeycomb ferromagnets}.--When the honeycomb magnet nanoribbon exhibits FM order ($J<0$), the Heisenberg Hamiltonian (Eq.~\ref{Heisenberg}) can be second-quantized by the Holstein-Primakoff transformation \cite{holstein1940} $S_A^+(\bm r) = (2S-a_{\bm r}^\dagger a_{\bm r})^{1/2}a_{\bm r}$ and $S_A^z(\bm r)=S-a_{\bm r}^\dagger a_{\bm r}$ [$S_B^+(\bm r) = (2S-b_{\bm r}^\dagger b_{\bm r})^{1/2}b_{\bm r}$ and $S_B^z(\bm r)=S-b_{\bm r}^\dagger b_{\bm r}$ ], where $a_{\bm r}$ ($b_{\bm r}$) is the magnon annihilation operator associated with the $A$ ($B$) sublattice. The resultant magnon tight-binding Hamiltonian to the bilinear order reads
\begin{equation} \label{FM_TB}
H^{\text{FM}}= \sum_{\bm r, i}  J_i S(a_{\bm r}^\dagger b_{\bm r+\bm \alpha_i}+a_{\bm r} b_{\bm r + \bm \alpha_i}^\dagger - a_{\bm r}^\dagger a_{\bm r} - b_{\bm r}^\dagger b_{\bm r}),
\end{equation}
where the FM ground state energy $E_G^{\text{FM}} = \sum_{\bm r,i}  J_iS^2$ has been deducted from the Heisenberg Hamiltonian. By applying Fourier transform of the basis $(a_{\bm r}, b_{\bm r})^T = N_{\text{uc}}^{-1/2} \sum_{\bm k} e^{i\bm k \cdot \bm r} (a_{\bm k}, b_{\bm k})^T$, where $N_{\text{uc}}$ is the number of unit cells, we obtain the Bloch Hamiltonian matrix  
\begin{equation} \label{FM_Bloch}
\pazocal{H}_{\bm k} = \sum_i J_iS [\cos(\bm k \cdot \bm \alpha_i) \sigma^x - \sin(\bm k \cdot \bm \alpha_i) \sigma^y -\sigma^0],
\end{equation}
where $\sigma^{x,y}$ and $\sigma^0$ are the Pauli matrices and the identity matrix defined in the basis $(a_{\bm k}, b_{\bm k})^T$. Without lattice deformation, the Bloch Hamiltonian dispersion $\epsilon_{\bm k}/|J|S = 3 \pm [3+2\cos(\sqrt 3 k_xa) + 4\cos(\frac{\sqrt 3}{2}k_xa) \cos(\frac{3}{2} k_ya)]^{1/2}$ exhibits two Dirac cones [Fig.~\ref{fig1}(b)] at the BZ corners $\bm k_W^\eta = (\eta \frac{4\pi}{3\sqrt 3 a},0)$ with valley index $\eta = \pm 1$. For a nanoribbon with a pair of zigzag edges along the $x$ direction [Fig.~\ref{fig1}(c)], the band $\epsilon_{\bm k}$ in the infinite system becomes a cluster of bands [Fig.~\ref{fig2}(a)].

For a fictitious lattice deformation that alters the nearest-neighbor interactions according to $J_1=J_2=J+ \delta J$ and $J_3=J$, where the variation $\delta J$ is a constant, an effective Dirac theory can be obtained by linearizing the Bloch Hamiltonian (Eq.~\ref{FM_Bloch}) in the vicinity of BZ corners as
\begin{multline} \label{FM_BZ_corners}
h_{\bm q} = \hbar v_x^\eta \bigg(q_x + \eta\frac{2\delta J}{3aJ} \bigg)\sigma^x + \hbar v_y^\eta q_y\sigma^y -(3JS+2\delta JS)\sigma^0,
\end{multline}
where the magnon velocity is $(v_x^\eta, v_y^\eta) = \tfrac{3JSa}{2\hbar}(-\eta, 1)$. We note that $\delta J$ has two effects. On the one hand, it shifts the two Dirac cones uniformly in the energy dimension by an amount of $-2\delta JS$. This effect is rather trivial and can be greatly suppressed by a Zeeman field $B_Z=-2\delta JS/g\mu_B$, where $\mu_B$ is the Bohr magneton. Therefore, we will neglect this effect in the following. On the other hand, $\delta J$ displaces the two Dirac cones \emph{oppositely} in the momentum dimension by $\eta\tfrac{2\delta J}{3aJ}$. Although an electric field can also alter the positions of magnon Dirac cones according to the Aharonov-Casher effect \cite{aharonov1984, suppl}, the two Dirac cones are always translated identically, implying that the valley-sensitive displacement should be interpreted as an emergent vector potential $\pazocal {\vec A} = \eta \tfrac{2\hbar}{3ea} \tfrac{\delta J}{J}\hat x$ that cannot be compensated by electric fields. 

We obtain $\pazocal{\vec A}$ by using a constant $\delta J$. However, we are interested in twist deformation as shown in Fig.~\ref{fig1}(a), where $\delta J=\delta J(y)$ depends on the $y$ coordinate. We argue that even in this case, the effect of $\delta J(y)$ can be treated as a vector potential $\pazocal{\vec A}(y)$ shifting Dirac cones and inducing Landau quantization, provided that $\delta J(y)$ varies slowly on the lattice scale. To substantiate our claim, we now present the results of our numerical simulations on the tight-binding Hamiltonian of a zigzag nanoribbon with exponentially decaying interactions adopted. We find that dispersive Dirac-Landau levels are induced on right (left) of valley $K$ ($K'$) [Fig.~\ref{fig2}(b)], reflecting the valley sensitivity of the emergent vector potential. These LLs are doubly degenerate due to the contribution from the upper ($y>0$) and lower ($y<0$) sections of the nanoribbon, respectively. To better resolve these LLs, we calculate the spectral function $A(\omega, k_x) = -\tfrac{1}{\pi} \sum_y \lim_{\delta \rightarrow 0}  \Im [\omega + i \delta - \pazocal{H}_{k_x}]^{-1}_{yy}$ in the bulk and on the edges of the nanoribbon. The bulk origin of these LLs is confirmed by the bulk spectral function, defined to include the contribution of the central $50\%$ lattice sites. We have read off the energies $\epsilon_{\text{LL}_n}$ of the first few LLs marked by the crosses in Fig.~\ref{fig2}(c) and find the sequence $\epsilon_{\text{LL}_n}-\epsilon_{\text{LL}_0}$ indeed exhibits the expected $\sqrt{n}$ dependence on the LL index $n$. We note that the dispersive Dirac-Landau levels only reside in the vicinity of Dirac cones. This observation is best demonstrated by the fact that the first three LLs ($n=0, \pm 1, \pm 2$) associated with each valley are not connected through the bulk. Instead, the edge spectral function [Fig.~\ref{fig2}(d)], which considers the $20\%$ lattice sites on the edges, reveals that LLs originating from different valleys are connected by edge states.

\begin{figure}
\includegraphics[width = 8.6cm]{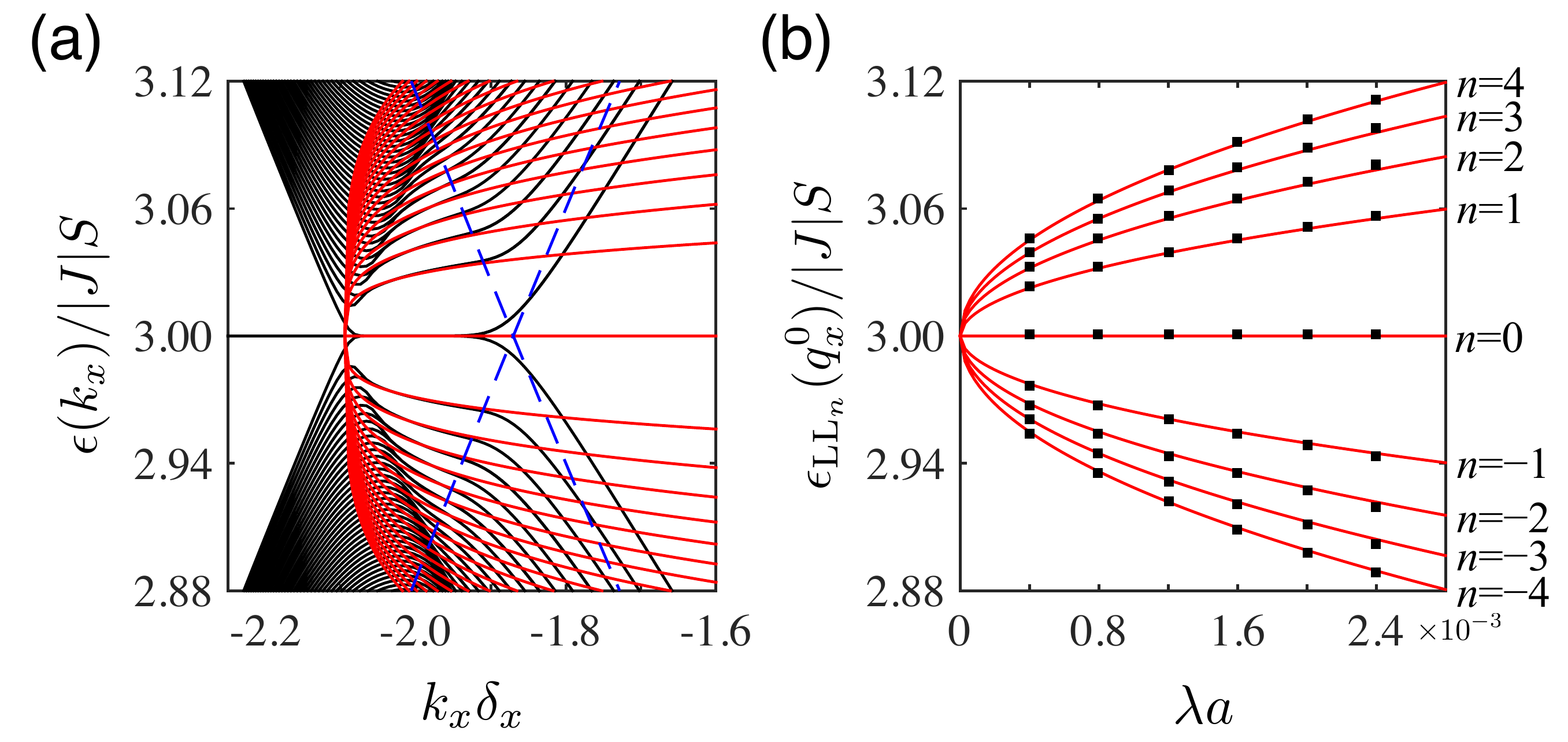}
\caption{Dispersive Dirac-Landau levels in a twisted FM nanoribbon.   (a) Numerically calculated magnon bands (black) for a nanoribbon with $2N=1200$ $A$ and $B$ sublattice sites and $\lambda a = 0.0008$. The theoretically predicted LLs (red) from Eq.~\ref{FM_LL} are overlaid. The blue dashed curves mark the position of the maximally displaced Dirac cone, which is $q_x^w \delta_x=0.225 $ on the right of valley $K$. (b) Dirac-Landau levels as a function of twist at a fixed momentum that is $q_x^0\delta_x = 0.139$ on the right of $K$ point. The theoretically predicted LLs (red) from Eq.~\ref{FM_LL} well fit the numerics (black dots) of the first few LLs. } \label{fig3}
\end{figure}

To obtain more insights on the twist-induced LLs, we now derive the dispersion of these LLs using quadratically decaying interactions in Eq.~\ref{quad}, which is a good approximation when the twist is sufficiently small.  A more generic derivation regarding large twist is given in the supplemental material (SM) \cite{suppl}.  Since the twist-induced vector potential $\pazocal{\vec A}(y) = \eta \frac{2\hbar}{3ea} \frac{\delta J(y)}{J} \hat x$ is to relocate the Dirac cones, for a specific momentum that is $q_x$ on the right of $K$ point, the Dirac point is shifted to this momentum by $\pazocal{A}_x=-\frac{\hbar}{e} q_x = -\frac{\hbar}{4ea} \lambda^2 y^2$, whose curl $\pazocal{B}_z = -\partial_y \pazocal{A}_x = \tfrac{\hbar}{2ea} \lambda^2 y = \tfrac{\hbar}{2ea} \lambda (4aq_x)^{1/2}$ results in the dispersive Dirac-Landau levels
\begin{equation} \label{FM_LL}
\begin{split}
\epsilon_{\text{LL}_n}(q_x) &= -3JS + \text{sgn}(n) \sqrt{\bigg| 2n \frac{e}{\hbar} \pazocal{B}_z \hbar v_x^\eta \hbar v_y^\eta \bigg|} 
\\
&= -3JS- \frac{3}{2}JS \sqrt{\lambda a} \sqrt[\leftroot{-3}\uproot{2}4]{4aq_x}\text{sgn}(n) \sqrt{|n|}.
\end{split}
\end{equation}
We apply numerical simulations on the tight-binding Hamiltonian of a zigzag FM nanoribbon with quadratically decaying interactions (Eq.~\ref{quad}) and find that Eq.~\ref{FM_LL} well captures the dispersion of the LLs as illustrated in Fig.~\ref{fig3}(a). For a fixed momentum slightly away from the Dirac point $K$, we also test the $\sqrt{\lambda a}$ dependence of the first few Dirac-Landau levels. These results are summarized in Fig.~\ref{fig3}(b). Far away from the valley $K$, where the LL wave functions overlap with the zigzag edges, LLs predicted in Eq.~\ref{FM_LL} begin to deviate from the numerically obtained bands, which are edge states rather than LLs. Considering the fact that LLs are pulled out by the displaced Dirac cone from the one at $K$, the width of the zeroth LL is then the maximal displacement $q_x^w=[-\tfrac{e}{\hbar} \pazocal{A}_x (y)]_{\text{max}} = \frac{\lambda^2W^2}{16a}$ of the Dirac cone and higher-order LLs have smaller widths because of the larger spatial extent of their wave functions. Therefore, Eq.~\ref{FM_LL} fits the numerics best between the Dirac cone at $K$ and the maximally displaced Dirac cone (blue dashed curves).

\begin{figure}
\includegraphics[width = 8.6cm]{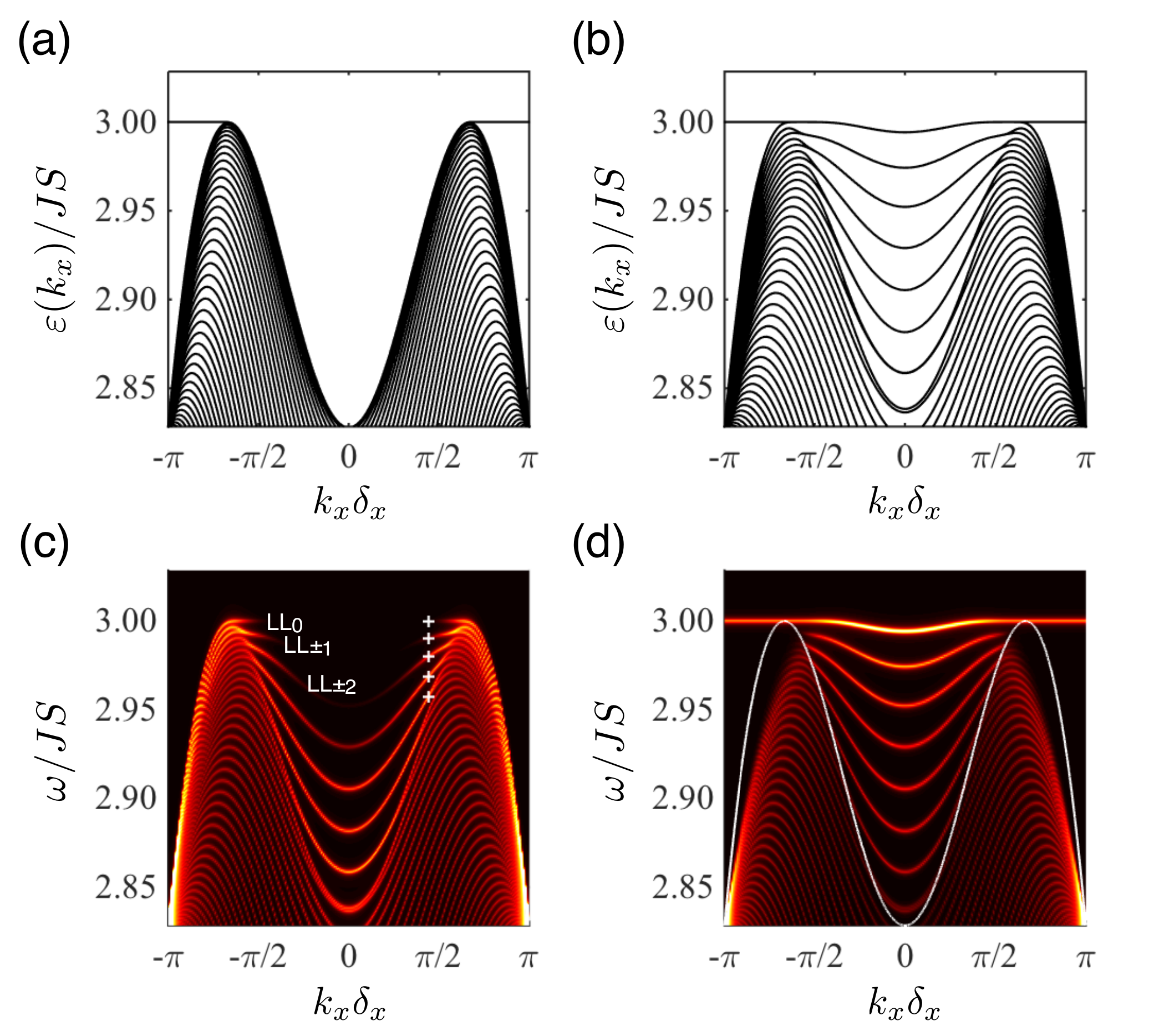}
\caption{Spectral properties of a zigzag AF nanoribbon with $2N=80$ $A$ and $B$ sublattice sites. (a) The spectrum of an untwisted nanoribbon exhibiting two quadratic peaks connected by flat edge states. (b) The spectrum of a twisted nanoribbon with $\lambda a =0.028$. (c) The bulk spectral function of (b) with dispersive equidistant LLs in the vicinity of quadratic peak $K$ labeled. The crosses mark the positions of LLs at the selected momentum $k_x\delta_x=1.382$. (d) The edge spectral function of (b) with white curves representing the envelopes of the quadratic peaks.}\label{fig4}
\end{figure}

\textit{Honeycomb antiferromagnets}.--When the honeycomb magnet exhibits AF order ($J>0$), the Heisenberg Hamiltonian (Eq.~\ref{Heisenberg}) can be second-quantized by the bipartite Holstein-Primakoff transformation $S_A^+(\bm r) = (2S-a_{\bm r}^\dagger a_{\bm r})^{1/2} a_{\bm r}$ and $S_A^z=S-a_{\bm r}^\dagger a_{\bm r}$ [$S_B^+(\bm r) = b_{\bm r}^\dagger(2S-b_{\bm r}^\dagger b_{\bm r})^{1/2}$ and $S_B^z=b_{\bm r}^\dagger b_{\bm r}-S$], where $a_{\bm r}$ ($b_{\bm r}$) is the magnon annihilation operator for spins on the $A$ ($B$) sublattice. The resultant magnon tight-binding model to the bilinear order reads
\begin{equation} \label{AF_TB}
H^{\text{AF}}=\sum_{\bm r, i} J_iS (a_{\bm r} b_{\bm r + \bm \alpha_i} + a_{\bm r}^\dagger b_{\bm r + \bm \alpha_i}^\dagger + a_{\bm r}^\dagger a_{\bm r} + b_{\bm r}^\dagger b_{\bm r}),
\end{equation}
where the N\'eel state energy $E_N = -\sum_{\bm r, i} J_iS^2$ has been deducted from the Heisenberg Hamiltonian. Applying Fourier transform of the basis $(a_{\bm r}, b_{\bm r})^T = N_{\text{uc}}^{-1/2}\sum_{\bm k} e^{i \bm k \cdot \bm r} (a_{\bm k}, b_{\bm k})^T$, we obtain the Bloch Hamiltonian 
\begin{equation} \label{AF_Bloch}
\mathcal{H}_{\bm k} = \sum_i J_iS [\cos(\bm k \cdot \bm \alpha_i)\tau^x -  \sin(\bm k \cdot \bm \alpha_i)\tau^y + \tau^0], 
\end{equation}
where $\tau^{x,y}$ and $\tau^0$ are Pauli matrices and identity matrix defined in basis $(a_{\bm k}, b_{-\bm k}^\dagger)^T$ and a constant term $-\sum_iJ_iS$ altering the N\'eel state energy is temporarily ignored \cite{suppl}. Without lattice deformation, the dispersion of $\mathcal{H}_{\bm k}$ can be obtained by Bogoliubov transformation \cite{suppl} as $\varepsilon_{\bm k}/JS= 3 \sqrt{1 - |\xi_{\bm k}|^2}$, which is doubly degenerate with $\xi_{\bm k} = \tfrac{1}{3} \sum_i e^{i \bm k \cdot \bm \alpha_i}$. $\varepsilon_{\bm k}$ exhibits two quadratic peaks at BZ corners $K/K'$ as illustrated by the blue curve in Fig.~\ref{fig1}(b). For a zigzag nanoribbon [Fig.~\ref{fig1}(c)], the band $\varepsilon_{\bm k}$ of the infinite system becomes a set of bands [Fig.~\ref{fig4}(a)].

For a fictitious lattice deformation that alters the nearest-neighbor interactions according to $J_1=J_2=J+\delta J$ and $J_3=J$ with a constant $\delta J$, the magnon dispersion in the vicinity of BZ corners is \cite{suppl}
\begin{multline} \label{AF_BZ_corners}
\varepsilon_{\bm q} =  (3JS + 2\delta JS) - \frac{3JS}{8}  a^2\bigg[\bigg(q_x + \eta \frac{2\delta J}{3aJ}\bigg)^2+q_y^2\bigg].
\end{multline}
Similar to the FM case, a constant $\delta J$ can shift the quadratic peaks identically in the energy dimension and oppositely in the momentum dimension. The former can be compensated by an external magnetic field, while the latter can still be interpreted as an emergent vector potential $\mathcal{\vec A} = \eta \tfrac{2\hbar}{3ea} \tfrac{\delta J}{J} \hat x$ same as that in ferromagnets. In the presence of twist deformation, $\delta J=\delta J(y)$ varies along the $y$ direction. But as long as $\delta J(y)$ varies slowly on the lattice scale, the effect of $\delta J(y)$ can still be treated as a vector potential $\mathcal{\vec A}(y)$ shifting the quadratic peaks and producing LLs.

To support our argument, we numerically calculate the band structure of a zigzag AF nanoribbon with exponentially decaying interactions employed. We find dispersive \emph{equidistant} LLs on the right (left) of quadratic peak $K$ ($K'$) [Fig.~\ref{fig4}(b)]. These LLs are 4-fold degenerate except for the zeroth LL, which is doubly degenerate. We have read off the LL energies $\varepsilon_{\text{LL}_n}$ of the first few LLs marked by the crosses in Fig.~\ref{fig4}(c) and find the sequence $\varepsilon_{\text{LL}_n} - \varepsilon_{\text{LL}_0}$ indeed shows the expected $n$ dependence on LL index $n$. This is consistent with the quadratic dispersion (Eq.~\ref{AF_BZ_corners}) at BZ corners. The bulk spectral function [Fig.~\ref{fig4}(c)] confirms the bulk origin of the LLs with best resolution for the first three LLs ($n=0, \pm 1, \pm 2$) of each quadratic peak, which are connected by edge states rather than through the bulk as illustrated by the edge spectral function [Fig.~\ref{fig4}(d)]. 

\begin{figure}
\includegraphics[width = 8.6cm]{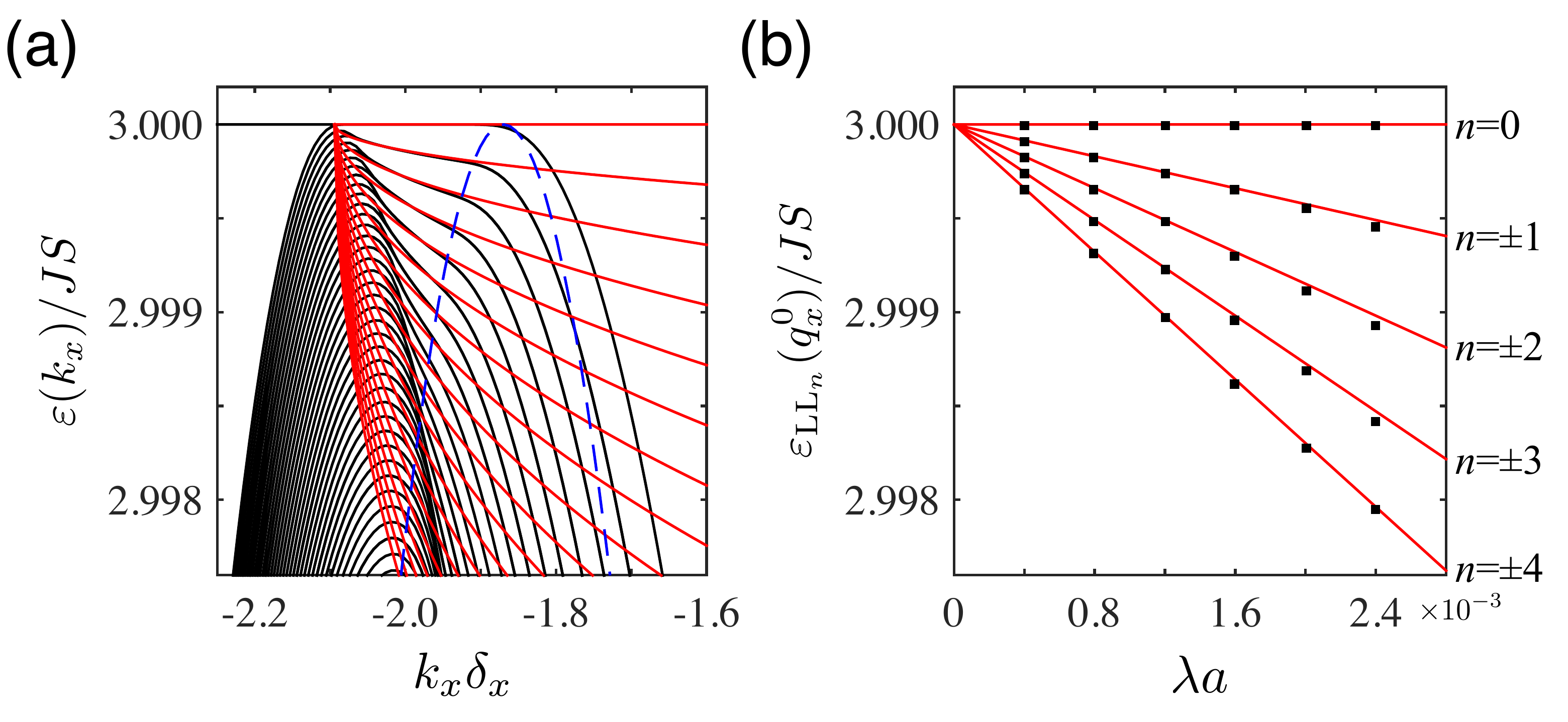}
\caption{Dispersive equidistant Landau levels in a twisted AF nanoribbon. (a) Numerically calculated band structure (black) for a nanoribbon with $2N=1200$ $A$ and $B$ sublattice sites and $\lambda a=0.0008$. The theoretically predicted LLs (red) from Eq.~\ref{AF_LL} are overlaid. The blue dashed curves mark the position of the maximally displaced quadratic peak, which is $q_x^w \delta_x = 0.225$ on the right of quadratic peak $K$. (b) Equidistant LLs as a function of twist at fixed momentum that is $q_x^0 \delta_x=0.139$ on the right of $K$ point. The theoretically predicted LLs (red) from Eq.~\ref{AF_LL} well fit the numerics (black dots) of the first few LLs.} \label{fig5}
\end{figure}

Following the technique we have developed for ferromagnets, we immediately find the momentum dependence of the twist-induced vector potential and the resulting gauge field as $\mathcal{A}_x= -\tfrac{\hbar}{e}q_x$ and  $\mathcal{B}_z = \tfrac{\hbar}{2ea} \lambda (4aq_x)^{1/2}$, respectively. The latter results in the dispersive equidistant LLs
\begin{equation} \label{AF_LL}
\begin{split}
\varepsilon_{\text{LL}_n}(q_x) &= \sqrt{(3JS)^2-\bigg| 2n\frac{e}{\hbar} \mathcal{B}_z \hbar v_x^\eta \hbar v_y^\eta \bigg|}
\\
& =3JS - \frac{3}{8}JS \lambda a \sqrt{4aq_x} |n|.
\end{split}
\end{equation}
We apply numerical simulations of the tight-binding Hamiltonian of a zigzag AF nanoribbon with quadratically decaying interactions (Eq.~\ref{fig2}). As illustrated in Fig.~\ref{fig5}(a), we find Eq.~\ref{AF_LL} well captures the LL dispersion between the quadratic peak at $K$ and the maximally displaced quadratic peak (dashed blue curve), which is $q_x^w=\tfrac{\lambda^2W^2}{16a}$ on the right of quadratic peak $K$. We also examine the linear dependence on $\lambda a$ of the first few LLs at a fixed momentum slightly away from quadratic peak $K$. These results are summarized in Fig.~\ref{fig5}(b).

\textit{Conclusions}.--We have studied the Landau level dispersion of twisted nanoribbons of ferromagnetic and antiferromagnetic honeycomb magnets. We elucidate that these Landau levels are pulled out by the twist-displaced magnon Dirac cones (quadratic peaks) from those located at Brillouin zone corners such that a correspondence can be constructed between the crystal momentum of the nanoribbon and the twist-induced gauge field, from which the dispersion of Landau levels can be explicitly derived for the ferromagnetic (antiferromagnetic) nanoribbons. Our proposal may be carried out with honeycomb ferromagnets CrX$_3$ (X=F, Cl, Br, I) \cite{huangbevin2017, pershoguba2018} and antiferromagnet MnPS$_3$ \cite{shiomi2017}. The required Zeeman field canceling the twist-induced onsite energy may be provided by a fine-tuned array of magnetic force microscope tips \cite{martin1987}, and the dispersive Landau levels can be imaged by neutron scattering \cite{brockhouse1957}. 

\begin{acknowledgments}
The authors are indebted to R. Moessner, M. Franz, and H. Kondo for insightful discussions. ZS is supported in part by project A02 of the CRC-TR 183.
\end{acknowledgments}

\bibliography{a1}

\begin{thebibliography}{46}
\expandafter\ifx\csname natexlab\endcsname\relax\def\natexlab#1{#1}\fi
\expandafter\ifx\csname bibnamefont\endcsname\relax
  \def\bibnamefont#1{#1}\fi
\expandafter\ifx\csname bibfnamefont\endcsname\relax
  \def\bibfnamefont#1{#1}\fi
\expandafter\ifx\csname citenamefont\endcsname\relax
  \def\citenamefont#1{#1}\fi
\expandafter\ifx\csname url\endcsname\relax
  \def\url#1{\texttt{#1}}\fi
\expandafter\ifx\csname urlprefix\endcsname\relax\def\urlprefix{URL }\fi
\providecommand{\bibinfo}[2]{#2}
\providecommand{\eprint}[2][]{\url{#2}}

\bibitem[{\citenamefont{Huang et~al.}(2017{\natexlab{a}})\citenamefont{Huang,
  Jin, Cui, Zhai, Mei, and Liu}}]{huangbing2017}
\bibinfo{author}{\bibfnamefont{B.}~\bibnamefont{Huang}},
  \bibinfo{author}{\bibfnamefont{K.-H.} \bibnamefont{Jin}},
  \bibinfo{author}{\bibfnamefont{B.}~\bibnamefont{Cui}},
  \bibinfo{author}{\bibfnamefont{F.}~\bibnamefont{Zhai}},
  \bibinfo{author}{\bibfnamefont{J.}~\bibnamefont{Mei}}, \bibnamefont{and}
  \bibinfo{author}{\bibfnamefont{F.}~\bibnamefont{Liu}}, \bibinfo{journal}{Nat.
  Commun.} \textbf{\bibinfo{volume}{8}}, \bibinfo{pages}{1}
  (\bibinfo{year}{2017}{\natexlab{a}}).

\bibitem[{\citenamefont{{\c{S}}ahin et~al.}(2019)\citenamefont{{\c{S}}ahin,
  Vignale, and Flatt{\'e}}}]{csahin2019}
\bibinfo{author}{\bibfnamefont{C.}~\bibnamefont{{\c{S}}ahin}},
  \bibinfo{author}{\bibfnamefont{G.}~\bibnamefont{Vignale}}, \bibnamefont{and}
  \bibinfo{author}{\bibfnamefont{M.~E.} \bibnamefont{Flatt{\'e}}},
  \bibinfo{journal}{Phys. Rev. Materials} \textbf{\bibinfo{volume}{3}},
  \bibinfo{pages}{014401} (\bibinfo{year}{2019}).

\bibitem[{\citenamefont{Meng et~al.}(2019)\citenamefont{Meng, Pandey, Jeong,
  Fu, Yang, Chen, Singh, He, Xu, Zhou et~al.}}]{meng2019}
\bibinfo{author}{\bibfnamefont{X.}~\bibnamefont{Meng}},
  \bibinfo{author}{\bibfnamefont{T.}~\bibnamefont{Pandey}},
  \bibinfo{author}{\bibfnamefont{J.}~\bibnamefont{Jeong}},
  \bibinfo{author}{\bibfnamefont{S.}~\bibnamefont{Fu}},
  \bibinfo{author}{\bibfnamefont{J.}~\bibnamefont{Yang}},
  \bibinfo{author}{\bibfnamefont{K.}~\bibnamefont{Chen}},
  \bibinfo{author}{\bibfnamefont{A.}~\bibnamefont{Singh}},
  \bibinfo{author}{\bibfnamefont{F.}~\bibnamefont{He}},
  \bibinfo{author}{\bibfnamefont{X.}~\bibnamefont{Xu}},
  \bibinfo{author}{\bibfnamefont{J.}~\bibnamefont{Zhou}}, \bibnamefont{et~al.},
  \bibinfo{journal}{Phys. Rev. Lett.} \textbf{\bibinfo{volume}{122}},
  \bibinfo{pages}{155901} (\bibinfo{year}{2019}).

\bibitem[{\citenamefont{Seijas-Bellido
  et~al.}(2019)\citenamefont{Seijas-Bellido, Rurali, {\'I}{\~n}iguez, Colombo,
  and Melis}}]{seijas2019}
\bibinfo{author}{\bibfnamefont{J.~A.} \bibnamefont{Seijas-Bellido}},
  \bibinfo{author}{\bibfnamefont{R.}~\bibnamefont{Rurali}},
  \bibinfo{author}{\bibfnamefont{J.}~\bibnamefont{{\'I}{\~n}iguez}},
  \bibinfo{author}{\bibfnamefont{L.}~\bibnamefont{Colombo}}, \bibnamefont{and}
  \bibinfo{author}{\bibfnamefont{C.}~\bibnamefont{Melis}},
  \bibinfo{journal}{Phys. Rev. Materials} \textbf{\bibinfo{volume}{3}},
  \bibinfo{pages}{065401} (\bibinfo{year}{2019}).

\bibitem[{\citenamefont{Cifuentes-Quintal
  et~al.}(2016)\citenamefont{Cifuentes-Quintal, de~la Pe{\~n}a-Seaman, Heid,
  de~Coss, and Bohnen}}]{cifuentes2016}
\bibinfo{author}{\bibfnamefont{M.}~\bibnamefont{Cifuentes-Quintal}},
  \bibinfo{author}{\bibfnamefont{O.}~\bibnamefont{de~la Pe{\~n}a-Seaman}},
  \bibinfo{author}{\bibfnamefont{R.}~\bibnamefont{Heid}},
  \bibinfo{author}{\bibfnamefont{R.}~\bibnamefont{de~Coss}}, \bibnamefont{and}
  \bibinfo{author}{\bibfnamefont{K.-P.} \bibnamefont{Bohnen}},
  \bibinfo{journal}{Phys. Rev. B} \textbf{\bibinfo{volume}{94}},
  \bibinfo{pages}{085401} (\bibinfo{year}{2016}).

\bibitem[{\citenamefont{Bistritzer and MacDonald}(2011)}]{bistritzer2011}
\bibinfo{author}{\bibfnamefont{R.}~\bibnamefont{Bistritzer}} \bibnamefont{and}
  \bibinfo{author}{\bibfnamefont{A.~H.} \bibnamefont{MacDonald}},
  \bibinfo{journal}{Proc. Natl. Acad. Sci. U.S.A.}
  \textbf{\bibinfo{volume}{108}}, \bibinfo{pages}{12233}
  (\bibinfo{year}{2011}).

\bibitem[{\citenamefont{Cao et~al.}(2018{\natexlab{a}})\citenamefont{Cao,
  Fatemi, Demir, Fang, Tomarken, Luo, Sanchez-Yamagishi, Watanabe, Taniguchi,
  Kaxiras et~al.}}]{cao2018a}
\bibinfo{author}{\bibfnamefont{Y.}~\bibnamefont{Cao}},
  \bibinfo{author}{\bibfnamefont{V.}~\bibnamefont{Fatemi}},
  \bibinfo{author}{\bibfnamefont{A.}~\bibnamefont{Demir}},
  \bibinfo{author}{\bibfnamefont{S.}~\bibnamefont{Fang}},
  \bibinfo{author}{\bibfnamefont{S.~L.} \bibnamefont{Tomarken}},
  \bibinfo{author}{\bibfnamefont{J.~Y.} \bibnamefont{Luo}},
  \bibinfo{author}{\bibfnamefont{J.~D.} \bibnamefont{Sanchez-Yamagishi}},
  \bibinfo{author}{\bibfnamefont{K.}~\bibnamefont{Watanabe}},
  \bibinfo{author}{\bibfnamefont{T.}~\bibnamefont{Taniguchi}},
  \bibinfo{author}{\bibfnamefont{E.}~\bibnamefont{Kaxiras}},
  \bibnamefont{et~al.}, \bibinfo{journal}{Nature}
  \textbf{\bibinfo{volume}{556}}, \bibinfo{pages}{80}
  (\bibinfo{year}{2018}{\natexlab{a}}).

\bibitem[{\citenamefont{Cao et~al.}(2018{\natexlab{b}})\citenamefont{Cao,
  Fatemi, Fang, Watanabe, Taniguchi, Kaxiras, and Jarillo-Herrero}}]{cao2018b}
\bibinfo{author}{\bibfnamefont{Y.}~\bibnamefont{Cao}},
  \bibinfo{author}{\bibfnamefont{V.}~\bibnamefont{Fatemi}},
  \bibinfo{author}{\bibfnamefont{S.}~\bibnamefont{Fang}},
  \bibinfo{author}{\bibfnamefont{K.}~\bibnamefont{Watanabe}},
  \bibinfo{author}{\bibfnamefont{T.}~\bibnamefont{Taniguchi}},
  \bibinfo{author}{\bibfnamefont{E.}~\bibnamefont{Kaxiras}}, \bibnamefont{and}
  \bibinfo{author}{\bibfnamefont{P.}~\bibnamefont{Jarillo-Herrero}},
  \bibinfo{journal}{Nature} \textbf{\bibinfo{volume}{556}}, \bibinfo{pages}{43}
  (\bibinfo{year}{2018}{\natexlab{b}}).

\bibitem[{\citenamefont{Zhu et~al.}(2016)\citenamefont{Zhu, Li, and
  Li}}]{zhu2016}
\bibinfo{author}{\bibfnamefont{Z.}~\bibnamefont{Zhu}},
  \bibinfo{author}{\bibfnamefont{M.}~\bibnamefont{Li}}, \bibnamefont{and}
  \bibinfo{author}{\bibfnamefont{J.}~\bibnamefont{Li}}, \bibinfo{journal}{Phys.
  Rev. B} \textbf{\bibinfo{volume}{94}}, \bibinfo{pages}{155121}
  (\bibinfo{year}{2016}).

\bibitem[{\citenamefont{Shao et~al.}(2017)\citenamefont{Shao, Ruan, Wu, Chen,
  Guo, Zhang, Sun, Sheng, and Xing}}]{shao2017}
\bibinfo{author}{\bibfnamefont{D.}~\bibnamefont{Shao}},
  \bibinfo{author}{\bibfnamefont{J.}~\bibnamefont{Ruan}},
  \bibinfo{author}{\bibfnamefont{J.}~\bibnamefont{Wu}},
  \bibinfo{author}{\bibfnamefont{T.}~\bibnamefont{Chen}},
  \bibinfo{author}{\bibfnamefont{Z.}~\bibnamefont{Guo}},
  \bibinfo{author}{\bibfnamefont{H.}~\bibnamefont{Zhang}},
  \bibinfo{author}{\bibfnamefont{J.}~\bibnamefont{Sun}},
  \bibinfo{author}{\bibfnamefont{L.}~\bibnamefont{Sheng}}, \bibnamefont{and}
  \bibinfo{author}{\bibfnamefont{D.}~\bibnamefont{Xing}},
  \bibinfo{journal}{Phys. Rev. B} \textbf{\bibinfo{volume}{96}},
  \bibinfo{pages}{075112} (\bibinfo{year}{2017}).

\bibitem[{\citenamefont{Guan et~al.}(2017)\citenamefont{Guan, Yu, Liu, Liu,
  Dong, Lu, Yao, and Yang}}]{guan2017}
\bibinfo{author}{\bibfnamefont{S.}~\bibnamefont{Guan}},
  \bibinfo{author}{\bibfnamefont{Z.-M.} \bibnamefont{Yu}},
  \bibinfo{author}{\bibfnamefont{Y.}~\bibnamefont{Liu}},
  \bibinfo{author}{\bibfnamefont{G.-B.} \bibnamefont{Liu}},
  \bibinfo{author}{\bibfnamefont{L.}~\bibnamefont{Dong}},
  \bibinfo{author}{\bibfnamefont{Y.}~\bibnamefont{Lu}},
  \bibinfo{author}{\bibfnamefont{Y.}~\bibnamefont{Yao}}, \bibnamefont{and}
  \bibinfo{author}{\bibfnamefont{S.~A.} \bibnamefont{Yang}},
  \bibinfo{journal}{Npj Quantum Mater.} \textbf{\bibinfo{volume}{2}},
  \bibinfo{pages}{1} (\bibinfo{year}{2017}).

\bibitem[{\citenamefont{Owerre}(2018)}]{owerre2018}
\bibinfo{author}{\bibfnamefont{S.}~\bibnamefont{Owerre}}, \bibinfo{journal}{J.
  Phys.: Condens. Matter} \textbf{\bibinfo{volume}{30}},
  \bibinfo{pages}{245803} (\bibinfo{year}{2018}).

\bibitem[{\citenamefont{Zhang et~al.}(2019)\citenamefont{Zhang, Luo, Chen, Zhu,
  Yu, Fang, and Weng}}]{zhang2019}
\bibinfo{author}{\bibfnamefont{W.}~\bibnamefont{Zhang}},
  \bibinfo{author}{\bibfnamefont{K.}~\bibnamefont{Luo}},
  \bibinfo{author}{\bibfnamefont{Z.}~\bibnamefont{Chen}},
  \bibinfo{author}{\bibfnamefont{Z.}~\bibnamefont{Zhu}},
  \bibinfo{author}{\bibfnamefont{R.}~\bibnamefont{Yu}},
  \bibinfo{author}{\bibfnamefont{C.}~\bibnamefont{Fang}}, \bibnamefont{and}
  \bibinfo{author}{\bibfnamefont{H.}~\bibnamefont{Weng}}, \bibinfo{journal}{Npj
  Comput. Mater.} \textbf{\bibinfo{volume}{5}}, \bibinfo{pages}{1}
  (\bibinfo{year}{2019}).

\bibitem[{\citenamefont{Mutch et~al.}(2019)\citenamefont{Mutch, Chen, Went,
  Qian, Wilson, Andreev, Chen, and Chu}}]{mutch2019}
\bibinfo{author}{\bibfnamefont{J.}~\bibnamefont{Mutch}},
  \bibinfo{author}{\bibfnamefont{W.-C.} \bibnamefont{Chen}},
  \bibinfo{author}{\bibfnamefont{P.}~\bibnamefont{Went}},
  \bibinfo{author}{\bibfnamefont{T.}~\bibnamefont{Qian}},
  \bibinfo{author}{\bibfnamefont{I.~Z.} \bibnamefont{Wilson}},
  \bibinfo{author}{\bibfnamefont{A.}~\bibnamefont{Andreev}},
  \bibinfo{author}{\bibfnamefont{C.-C.} \bibnamefont{Chen}}, \bibnamefont{and}
  \bibinfo{author}{\bibfnamefont{J.-H.} \bibnamefont{Chu}},
  \bibinfo{journal}{Sci. Adv.} \textbf{\bibinfo{volume}{5}},
  \bibinfo{pages}{eaav9771} (\bibinfo{year}{2019}).

\bibitem[{\citenamefont{Guinea et~al.}(2010{\natexlab{a}})\citenamefont{Guinea,
  Katsnelson, and Geim}}]{guinea2010a}
\bibinfo{author}{\bibfnamefont{F.}~\bibnamefont{Guinea}},
  \bibinfo{author}{\bibfnamefont{M.}~\bibnamefont{Katsnelson}},
  \bibnamefont{and} \bibinfo{author}{\bibfnamefont{A.}~\bibnamefont{Geim}},
  \bibinfo{journal}{Nat. Phys.} \textbf{\bibinfo{volume}{6}},
  \bibinfo{pages}{30} (\bibinfo{year}{2010}{\natexlab{a}}).

\bibitem[{\citenamefont{Levy et~al.}(2010)\citenamefont{Levy, Burke, Meaker,
  Panlasigui, Zettl, Guinea, Neto, and Crommie}}]{levy2010}
\bibinfo{author}{\bibfnamefont{N.}~\bibnamefont{Levy}},
  \bibinfo{author}{\bibfnamefont{S.}~\bibnamefont{Burke}},
  \bibinfo{author}{\bibfnamefont{K.}~\bibnamefont{Meaker}},
  \bibinfo{author}{\bibfnamefont{M.}~\bibnamefont{Panlasigui}},
  \bibinfo{author}{\bibfnamefont{A.}~\bibnamefont{Zettl}},
  \bibinfo{author}{\bibfnamefont{F.}~\bibnamefont{Guinea}},
  \bibinfo{author}{\bibfnamefont{A.~C.} \bibnamefont{Neto}}, \bibnamefont{and}
  \bibinfo{author}{\bibfnamefont{M.}~\bibnamefont{Crommie}},
  \bibinfo{journal}{Science} \textbf{\bibinfo{volume}{329}},
  \bibinfo{pages}{544} (\bibinfo{year}{2010}).

\bibitem[{\citenamefont{Vozmediano et~al.}(2010)\citenamefont{Vozmediano,
  Katsnelson, and Guinea}}]{vozmediano2010}
\bibinfo{author}{\bibfnamefont{M.~A.} \bibnamefont{Vozmediano}},
  \bibinfo{author}{\bibfnamefont{M.}~\bibnamefont{Katsnelson}},
  \bibnamefont{and} \bibinfo{author}{\bibfnamefont{F.}~\bibnamefont{Guinea}},
  \bibinfo{journal}{Phys. Rep.} \textbf{\bibinfo{volume}{496}},
  \bibinfo{pages}{109} (\bibinfo{year}{2010}).

\bibitem[{\citenamefont{Rechtsman et~al.}(2013)\citenamefont{Rechtsman, Zeuner,
  T{\"u}nnermann, Nolte, Segev, and Szameit}}]{rechtsman2013}
\bibinfo{author}{\bibfnamefont{M.~C.} \bibnamefont{Rechtsman}},
  \bibinfo{author}{\bibfnamefont{J.~M.} \bibnamefont{Zeuner}},
  \bibinfo{author}{\bibfnamefont{A.}~\bibnamefont{T{\"u}nnermann}},
  \bibinfo{author}{\bibfnamefont{S.}~\bibnamefont{Nolte}},
  \bibinfo{author}{\bibfnamefont{M.}~\bibnamefont{Segev}}, \bibnamefont{and}
  \bibinfo{author}{\bibfnamefont{A.}~\bibnamefont{Szameit}},
  \bibinfo{journal}{Nat. Photonics} \textbf{\bibinfo{volume}{7}},
  \bibinfo{pages}{153} (\bibinfo{year}{2013}).

\bibitem[{\citenamefont{Cortijo et~al.}(2015)\citenamefont{Cortijo,
  Ferreir{\'o}s, Landsteiner, and Vozmediano}}]{cortijo2015}
\bibinfo{author}{\bibfnamefont{A.}~\bibnamefont{Cortijo}},
  \bibinfo{author}{\bibfnamefont{Y.}~\bibnamefont{Ferreir{\'o}s}},
  \bibinfo{author}{\bibfnamefont{K.}~\bibnamefont{Landsteiner}},
  \bibnamefont{and} \bibinfo{author}{\bibfnamefont{M.~A.}
  \bibnamefont{Vozmediano}}, \bibinfo{journal}{Phys. Rev. Lett.}
  \textbf{\bibinfo{volume}{115}}, \bibinfo{pages}{177202}
  (\bibinfo{year}{2015}).

\bibitem[{\citenamefont{Pikulin et~al.}(2016)\citenamefont{Pikulin, Chen, and
  Franz}}]{pikulin2016}
\bibinfo{author}{\bibfnamefont{D.}~\bibnamefont{Pikulin}},
  \bibinfo{author}{\bibfnamefont{A.}~\bibnamefont{Chen}}, \bibnamefont{and}
  \bibinfo{author}{\bibfnamefont{M.}~\bibnamefont{Franz}},
  \bibinfo{journal}{Phys. Rev. X} \textbf{\bibinfo{volume}{6}},
  \bibinfo{pages}{041021} (\bibinfo{year}{2016}).

\bibitem[{\citenamefont{Grushin et~al.}(2016)\citenamefont{Grushin, Venderbos,
  Vishwanath, and Ilan}}]{grushin2016}
\bibinfo{author}{\bibfnamefont{A.~G.} \bibnamefont{Grushin}},
  \bibinfo{author}{\bibfnamefont{J.~W.} \bibnamefont{Venderbos}},
  \bibinfo{author}{\bibfnamefont{A.}~\bibnamefont{Vishwanath}},
  \bibnamefont{and} \bibinfo{author}{\bibfnamefont{R.}~\bibnamefont{Ilan}},
  \bibinfo{journal}{Phys. Rev. X} \textbf{\bibinfo{volume}{6}},
  \bibinfo{pages}{041046} (\bibinfo{year}{2016}).

\bibitem[{\citenamefont{Cortijo et~al.}(2016)\citenamefont{Cortijo, Kharzeev,
  Landsteiner, and Vozmediano}}]{cortijo2016}
\bibinfo{author}{\bibfnamefont{A.}~\bibnamefont{Cortijo}},
  \bibinfo{author}{\bibfnamefont{D.}~\bibnamefont{Kharzeev}},
  \bibinfo{author}{\bibfnamefont{K.}~\bibnamefont{Landsteiner}},
  \bibnamefont{and} \bibinfo{author}{\bibfnamefont{M.~A.}
  \bibnamefont{Vozmediano}}, \bibinfo{journal}{Phys. Rev. B}
  \textbf{\bibinfo{volume}{94}}, \bibinfo{pages}{241405}
  (\bibinfo{year}{2016}).

\bibitem[{\citenamefont{Sumiyoshi and Fujimoto}(2016)}]{sumiyoshi2016}
\bibinfo{author}{\bibfnamefont{H.}~\bibnamefont{Sumiyoshi}} \bibnamefont{and}
  \bibinfo{author}{\bibfnamefont{S.}~\bibnamefont{Fujimoto}},
  \bibinfo{journal}{Phys. Rev. Lett.} \textbf{\bibinfo{volume}{116}},
  \bibinfo{pages}{166601} (\bibinfo{year}{2016}).

\bibitem[{\citenamefont{Arjona et~al.}(2017)\citenamefont{Arjona, Castro, and
  Vozmediano}}]{arjona2017}
\bibinfo{author}{\bibfnamefont{V.}~\bibnamefont{Arjona}},
  \bibinfo{author}{\bibfnamefont{E.~V.} \bibnamefont{Castro}},
  \bibnamefont{and} \bibinfo{author}{\bibfnamefont{M.~A.}
  \bibnamefont{Vozmediano}}, \bibinfo{journal}{Phys. Rev. B}
  \textbf{\bibinfo{volume}{96}}, \bibinfo{pages}{081110}
  (\bibinfo{year}{2017}).

\bibitem[{\citenamefont{Liu et~al.}(2017{\natexlab{a}})\citenamefont{Liu,
  Pikulin, and Franz}}]{liu2017a}
\bibinfo{author}{\bibfnamefont{T.}~\bibnamefont{Liu}},
  \bibinfo{author}{\bibfnamefont{D.}~\bibnamefont{Pikulin}}, \bibnamefont{and}
  \bibinfo{author}{\bibfnamefont{M.}~\bibnamefont{Franz}},
  \bibinfo{journal}{Phys. Rev. B} \textbf{\bibinfo{volume}{95}},
  \bibinfo{pages}{041201} (\bibinfo{year}{2017}{\natexlab{a}}).

\bibitem[{\citenamefont{Liu et~al.}(2017{\natexlab{b}})\citenamefont{Liu,
  Franz, and Fujimoto}}]{liu2017b}
\bibinfo{author}{\bibfnamefont{T.}~\bibnamefont{Liu}},
  \bibinfo{author}{\bibfnamefont{M.}~\bibnamefont{Franz}}, \bibnamefont{and}
  \bibinfo{author}{\bibfnamefont{S.}~\bibnamefont{Fujimoto}},
  \bibinfo{journal}{Phys. Rev. B} \textbf{\bibinfo{volume}{96}},
  \bibinfo{pages}{224518} (\bibinfo{year}{2017}{\natexlab{b}}).

\bibitem[{\citenamefont{Massarelli et~al.}(2017)\citenamefont{Massarelli,
  Wachtel, Wei, and Paramekanti}}]{massarelli2017}
\bibinfo{author}{\bibfnamefont{G.}~\bibnamefont{Massarelli}},
  \bibinfo{author}{\bibfnamefont{G.}~\bibnamefont{Wachtel}},
  \bibinfo{author}{\bibfnamefont{J.~Y.} \bibnamefont{Wei}}, \bibnamefont{and}
  \bibinfo{author}{\bibfnamefont{A.}~\bibnamefont{Paramekanti}},
  \bibinfo{journal}{Phys. Rev. B} \textbf{\bibinfo{volume}{96}},
  \bibinfo{pages}{224516} (\bibinfo{year}{2017}).

\bibitem[{\citenamefont{Matsushita et~al.}(2018)\citenamefont{Matsushita, Liu,
  Mizushima, and Fujimoto}}]{matsushita2018}
\bibinfo{author}{\bibfnamefont{T.}~\bibnamefont{Matsushita}},
  \bibinfo{author}{\bibfnamefont{T.}~\bibnamefont{Liu}},
  \bibinfo{author}{\bibfnamefont{T.}~\bibnamefont{Mizushima}},
  \bibnamefont{and} \bibinfo{author}{\bibfnamefont{S.}~\bibnamefont{Fujimoto}},
  \bibinfo{journal}{Phys. Rev. B} \textbf{\bibinfo{volume}{97}},
  \bibinfo{pages}{134519} (\bibinfo{year}{2018}).

\bibitem[{\citenamefont{Kobayashi et~al.}(2018)\citenamefont{Kobayashi,
  Matsushita, Mizushima, Tsuruta, and Fujimoto}}]{kobayashi2018}
\bibinfo{author}{\bibfnamefont{T.}~\bibnamefont{Kobayashi}},
  \bibinfo{author}{\bibfnamefont{T.}~\bibnamefont{Matsushita}},
  \bibinfo{author}{\bibfnamefont{T.}~\bibnamefont{Mizushima}},
  \bibinfo{author}{\bibfnamefont{A.}~\bibnamefont{Tsuruta}}, \bibnamefont{and}
  \bibinfo{author}{\bibfnamefont{S.}~\bibnamefont{Fujimoto}},
  \bibinfo{journal}{Phys. Rev. Lett.} \textbf{\bibinfo{volume}{121}},
  \bibinfo{pages}{207002} (\bibinfo{year}{2018}).

\bibitem[{\citenamefont{Nica and Franz}(2018)}]{nica2018}
\bibinfo{author}{\bibfnamefont{E.~M.} \bibnamefont{Nica}} \bibnamefont{and}
  \bibinfo{author}{\bibfnamefont{M.}~\bibnamefont{Franz}},
  \bibinfo{journal}{Phys. Rev. B} \textbf{\bibinfo{volume}{97}},
  \bibinfo{pages}{024520} (\bibinfo{year}{2018}).

\bibitem[{\citenamefont{Ferreiros and Vozmediano}(2018)}]{ferreiros2018}
\bibinfo{author}{\bibfnamefont{Y.}~\bibnamefont{Ferreiros}} \bibnamefont{and}
  \bibinfo{author}{\bibfnamefont{M.~A.} \bibnamefont{Vozmediano}},
  \bibinfo{journal}{Phys. Rev. B} \textbf{\bibinfo{volume}{97}},
  \bibinfo{pages}{054404} (\bibinfo{year}{2018}).

\bibitem[{\citenamefont{Liu and Shi}(2019)}]{liu2019}
\bibinfo{author}{\bibfnamefont{T.}~\bibnamefont{Liu}} \bibnamefont{and}
  \bibinfo{author}{\bibfnamefont{Z.}~\bibnamefont{Shi}},
  \bibinfo{journal}{Phys. Rev. B} \textbf{\bibinfo{volume}{99}},
  \bibinfo{pages}{214413} (\bibinfo{year}{2019}).

\bibitem[{\citenamefont{Liu}(2020)}]{liu2020}
\bibinfo{author}{\bibfnamefont{T.}~\bibnamefont{Liu}} (\bibinfo{year}{2020}),
  \eprint{2002.09289}.

\bibitem[{\citenamefont{Guinea et~al.}(2010{\natexlab{b}})\citenamefont{Guinea,
  Geim, Katsnelson, and Novoselov}}]{guinea2010b}
\bibinfo{author}{\bibfnamefont{F.}~\bibnamefont{Guinea}},
  \bibinfo{author}{\bibfnamefont{A.}~\bibnamefont{Geim}},
  \bibinfo{author}{\bibfnamefont{M.}~\bibnamefont{Katsnelson}},
  \bibnamefont{and}
  \bibinfo{author}{\bibfnamefont{K.}~\bibnamefont{Novoselov}},
  \bibinfo{journal}{Phys. Rev. B} \textbf{\bibinfo{volume}{81}},
  \bibinfo{pages}{035408} (\bibinfo{year}{2010}{\natexlab{b}}).

\bibitem[{\citenamefont{Chang et~al.}(2012)\citenamefont{Chang, Albash, and
  Haas}}]{chang2012}
\bibinfo{author}{\bibfnamefont{Y.}~\bibnamefont{Chang}},
  \bibinfo{author}{\bibfnamefont{T.}~\bibnamefont{Albash}}, \bibnamefont{and}
  \bibinfo{author}{\bibfnamefont{S.}~\bibnamefont{Haas}},
  \bibinfo{journal}{Phys. Rev. B} \textbf{\bibinfo{volume}{86}},
  \bibinfo{pages}{125402} (\bibinfo{year}{2012}).

\bibitem[{\citenamefont{Stuij et~al.}(2015)\citenamefont{Stuij, Jacobse,
  Juri{\v{c}}i{\'c}, and Smith}}]{stuij2015}
\bibinfo{author}{\bibfnamefont{S.}~\bibnamefont{Stuij}},
  \bibinfo{author}{\bibfnamefont{P.}~\bibnamefont{Jacobse}},
  \bibinfo{author}{\bibfnamefont{V.}~\bibnamefont{Juri{\v{c}}i{\'c}}},
  \bibnamefont{and} \bibinfo{author}{\bibfnamefont{C.~M.} \bibnamefont{Smith}},
  \bibinfo{journal}{Phys. Rev. B} \textbf{\bibinfo{volume}{92}},
  \bibinfo{pages}{075424} (\bibinfo{year}{2015}).

\bibitem[{\citenamefont{Zhang et~al.}(2014)\citenamefont{Zhang, Seifert, and
  Chang}}]{zhang2014}
\bibinfo{author}{\bibfnamefont{D.-B.} \bibnamefont{Zhang}},
  \bibinfo{author}{\bibfnamefont{G.}~\bibnamefont{Seifert}}, \bibnamefont{and}
  \bibinfo{author}{\bibfnamefont{K.}~\bibnamefont{Chang}},
  \bibinfo{journal}{Phys. Rev. Lett.} \textbf{\bibinfo{volume}{112}},
  \bibinfo{pages}{096805} (\bibinfo{year}{2014}).

\bibitem[{\citenamefont{Nayga et~al.}(2019)\citenamefont{Nayga, Rachel, and
  Vojta}}]{nayga2019}
\bibinfo{author}{\bibfnamefont{M.~M.} \bibnamefont{Nayga}},
  \bibinfo{author}{\bibfnamefont{S.}~\bibnamefont{Rachel}}, \bibnamefont{and}
  \bibinfo{author}{\bibfnamefont{M.}~\bibnamefont{Vojta}},
  \bibinfo{journal}{Phys. Rev. Lett.} \textbf{\bibinfo{volume}{123}},
  \bibinfo{pages}{207204} (\bibinfo{year}{2019}).

\bibitem[{\citenamefont{Holstein and Primakoff}(1940)}]{holstein1940}
\bibinfo{author}{\bibfnamefont{T.}~\bibnamefont{Holstein}} \bibnamefont{and}
  \bibinfo{author}{\bibfnamefont{H.}~\bibnamefont{Primakoff}},
  \bibinfo{journal}{Phys. Rev.} \textbf{\bibinfo{volume}{58}},
  \bibinfo{pages}{1098} (\bibinfo{year}{1940}).

\bibitem[{\citenamefont{Aharonov and Casher}(1984)}]{aharonov1984}
\bibinfo{author}{\bibfnamefont{Y.}~\bibnamefont{Aharonov}} \bibnamefont{and}
  \bibinfo{author}{\bibfnamefont{A.}~\bibnamefont{Casher}},
  \bibinfo{journal}{Phys. Rev. Lett.} \textbf{\bibinfo{volume}{53}},
  \bibinfo{pages}{319} (\bibinfo{year}{1984}).

\bibitem[{sup()}]{suppl}
\bibinfo{note}{See Supplemental Material for details.}

\bibitem[{\citenamefont{Huang et~al.}(2017{\natexlab{b}})\citenamefont{Huang,
  Clark, Navarro-Moratalla, Klein, Cheng, Seyler, Zhong, Schmidgall, McGuire,
  Cobden et~al.}}]{huangbevin2017}
\bibinfo{author}{\bibfnamefont{B.}~\bibnamefont{Huang}},
  \bibinfo{author}{\bibfnamefont{G.}~\bibnamefont{Clark}},
  \bibinfo{author}{\bibfnamefont{E.}~\bibnamefont{Navarro-Moratalla}},
  \bibinfo{author}{\bibfnamefont{D.~R.} \bibnamefont{Klein}},
  \bibinfo{author}{\bibfnamefont{R.}~\bibnamefont{Cheng}},
  \bibinfo{author}{\bibfnamefont{K.~L.} \bibnamefont{Seyler}},
  \bibinfo{author}{\bibfnamefont{D.}~\bibnamefont{Zhong}},
  \bibinfo{author}{\bibfnamefont{E.}~\bibnamefont{Schmidgall}},
  \bibinfo{author}{\bibfnamefont{M.~A.} \bibnamefont{McGuire}},
  \bibinfo{author}{\bibfnamefont{D.~H.} \bibnamefont{Cobden}},
  \bibnamefont{et~al.}, \bibinfo{journal}{Nature}
  \textbf{\bibinfo{volume}{546}}, \bibinfo{pages}{270}
  (\bibinfo{year}{2017}{\natexlab{b}}).

\bibitem[{\citenamefont{Pershoguba et~al.}(2018)\citenamefont{Pershoguba,
  Banerjee, Lashley, Park, {\AA}gren, Aeppli, and Balatsky}}]{pershoguba2018}
\bibinfo{author}{\bibfnamefont{S.~S.} \bibnamefont{Pershoguba}},
  \bibinfo{author}{\bibfnamefont{S.}~\bibnamefont{Banerjee}},
  \bibinfo{author}{\bibfnamefont{J.}~\bibnamefont{Lashley}},
  \bibinfo{author}{\bibfnamefont{J.}~\bibnamefont{Park}},
  \bibinfo{author}{\bibfnamefont{H.}~\bibnamefont{{\AA}gren}},
  \bibinfo{author}{\bibfnamefont{G.}~\bibnamefont{Aeppli}}, \bibnamefont{and}
  \bibinfo{author}{\bibfnamefont{A.~V.} \bibnamefont{Balatsky}},
  \bibinfo{journal}{Phys. Rev. X} \textbf{\bibinfo{volume}{8}},
  \bibinfo{pages}{011010} (\bibinfo{year}{2018}).

\bibitem[{\citenamefont{Shiomi et~al.}(2017)\citenamefont{Shiomi, Takashima,
  and Saitoh}}]{shiomi2017}
\bibinfo{author}{\bibfnamefont{Y.}~\bibnamefont{Shiomi}},
  \bibinfo{author}{\bibfnamefont{R.}~\bibnamefont{Takashima}},
  \bibnamefont{and} \bibinfo{author}{\bibfnamefont{E.}~\bibnamefont{Saitoh}},
  \bibinfo{journal}{Phys. Rev. B} \textbf{\bibinfo{volume}{96}},
  \bibinfo{pages}{134425} (\bibinfo{year}{2017}).

\bibitem[{\citenamefont{Martin and Wickramasinghe}(1987)}]{martin1987}
\bibinfo{author}{\bibfnamefont{Y.}~\bibnamefont{Martin}} \bibnamefont{and}
  \bibinfo{author}{\bibfnamefont{H.~K.} \bibnamefont{Wickramasinghe}},
  \bibinfo{journal}{Appl. Phys. Lett.} \textbf{\bibinfo{volume}{50}},
  \bibinfo{pages}{1455} (\bibinfo{year}{1987}).

\bibitem[{\citenamefont{Brockhouse}(1957)}]{brockhouse1957}
\bibinfo{author}{\bibfnamefont{B.}~\bibnamefont{Brockhouse}},
  \bibinfo{journal}{Phys. Rev.} \textbf{\bibinfo{volume}{106}},
  \bibinfo{pages}{859} (\bibinfo{year}{1957}).

\end{thebibliography}

%
%
%
%
%
%


\end{document}